\documentclass[twocolumn,prb,showpacs,multicol,amsmath,amssymb]{revtex4}
\usepackage[dvips]{graphicx}

\newcommand{\be}{\begin{equation}}
\newcommand{\ee}{\end{equation}}

\newcommand{\bea}{\begin{eqnarray}}
\newcommand{\eea}{\end{eqnarray}}
\newcommand{\bd}{\begin{displaymath}}
\newcommand{\ed}{\end{displaymath}}
\newcommand{\ba}{\begin{array}}
\newcommand{\ea}{\end{array}}
\newcommand{\bi}{\begin{itemize}}
\newcommand{\ei}{\end{itemize}}
\newcommand{\bc}{\begin{center}}
\newcommand{\ec}{\end{center}}
\newcommand{\bfl}{\begin{flushleft}}
\newcommand{\efl}{\end{flushleft}}
\newcommand{\bfr}{\begin{flushright}}
\newcommand{\efr}{\end{flushright}}

\def\6{\partial}

\def\no{\nonumber \\}

\def\={\!\!\!&=&\!\!\!}
\def\+{\!\!\!&&\!\!\!+~}
\def\-{\!\!\!&&\!\!\!-~}

\begin{document}
%\date{23 October  2007}
\title{Heat capacity of Schottky type in low-dimensional spin system}

\author {Saeed Mahdavifar$^{1,3}$ and Alireza Akbari$^{2,3}$}

\affiliation{ $^{1}$Department of Physics, Guilan University,
P.O.Box 41335-1914, Rasht, Iran\\
$^{2}$Max-Planck-Institut f\"ur Physik komplexer Systeme, N\"othnitzer Str.38,
01187 Dresden, Germany\\
$^{3}$Institute for Advanced Studies in Basic Sciences, P.O.Box
45195-1159, Zanjan, Iran}

\begin{abstract}
The heat capacity of low-dimensional spin systems is studied using
theoretical and numerical techniques. Keeping only two energy
states, the system is mapped onto the two -level-system (TLS)
model. Using the low temperature Lanczos method, it is
confirmed that the behavior of $T_{M}$ and the energy gap as
functions of the control parameter is the same in the two models
studied; a conclusion that can probably be extrapolated to the
general case of any system that possesses an energy gap.

\end{abstract}


\maketitle

%%%%%%%%%%%%%%%%%%%%%%%%%%%%%%%%%%%%%%%%%%%%%%%%%%%%%%%%%%%%%%%%%%%%%%%%%%
%%%%%%%%%%%%%%%%%%%%%%%%%%     Section I      %%%%%%%%%%%%%%%%%%%%%%%%%%%%
%%%%%%%%%%%%%%%%%%%%%%%%%%%%%%%%%%%%%%%%%%%%%%%%%%%%%%%%%%%%%%%%%%%%%%%%%%
\section{Introduction}

Gaps in the energy spectrum play a crucial role in condensed matter physics.
 Fundamental properties in superconductivity or in the fractional quantum Hall
 effect originate from the existence of a gap between the ground
  state and the excited states. In particular, low dimensional quantum
  spin systems are extremely interesting to study the behavior of the gap.
  Many exact and numerical results on the one dimensional quantum spin systems
  with nearest neighbor couplings have been accumulated during last decades.

The 1D spin-1/2 system has been solved by Bethe\cite{bethe31} in
1931 with his famous ansatz. The ansatz allows the computation of
the energy eigenvalues. The isotropic 1D spin-1/2 system with
nearest neighbor couplings is gapless. The anisotropic spin-1/2
chain is denoted by the XXZ model\cite{kurmann82, dmitriev02}. The
Hamiltonian of the XXZ model on a periodic chain of $N$ sites is
%***********************************************************
\begin{eqnarray}
H=J\sum_{i=1}^{N}S_{i}^{x}S_{i+1}^{x}+S_{i}^{y}S_{i+1}^{y}+\Delta
S_{i}^{z}S_{i+1}^{z}, \label{hamiltoni}
\end{eqnarray}
%***********************************************************
where $J>0$ is the exchange coupling in the $xy$ easy plane,
$\Delta$ is the anisotropy in the $z$ direction.
 The Ising regime
is governed by $\Delta>1$ and there is a gap in the excitation
spectrum, while for $\Delta\leq -1$, the ground state is in the
ferromagnetic phase and there is a gap over the ferromagnetic
state. In the region $-1<\Delta\leq 1$, the ground state of the
system, is in the gapless spin-fluid phase.

The spin-1 system is not solvable with the Bethe ansatz or similar
techniques. The anisotropic spin-1 chain is only
gapless\cite{chen03} in the region $-1<\Delta<0$.
 Haldane\cite{haldane83} formulated in 1983 his famous conjecture that quantum spin
 chains (isotropic) with integer spin $S=1, 2, ..$ have a gap, where as chains with
 half-integer spin $S=1/2, 3/2, ...$ are gapless.

During the last two decades ladder-systems\cite{dagotto96} as quantum spin systems
between one and two dimensions have been studied as well with numerical methods.
 Concerning the ground state properties has been found that, ladder-systems with
 an even number of legs ($l=2, 4, 6, ...$) have a gap; those with an odd number
 ($l=1, 3, 5, ...$) do not. In particular, since the antiferromagnetic two-leg
 ladder systems have a gap in the spin excitation spectrum, they reveal extremely
 rich quantum behavior in the presence of a magnetic field\cite{schmidt05}.
 Such quantum phase transitions in spin systems with gapped excitation spectrum
 were indeed studied
 experimentally\cite{chaboussant97, chaboussant98-1, chaboussant98-2, arcon99, mayaffre00, watson01}.

On the other hand, investigating the behavior of the energy gap of
spin systems in vicinity of quantum critical points has attracted
much interest recently\cite{dmitriev02, saeed1, saeed2, saeed3}.
In general, the critical point of an thermodynamic system in the
Hamiltonian formulation is defined as the value at which the
energy gap vanishes as a power low, which is known as the scaling
behavior. The opening of the energy gap in the vicinity of the
quantum critical point is found to scale with a critical exponent.
The value of the critical gap exponent is very important to find
the universality class of a continuous quantum phase transition.

The discovery of gapless or gaped excitations have led to the
investigation of the thermodynamic properties. One of the most
important thermodynamic functions is the heat capacity. Usually,
there is a lambda-type anomaly in figure of the heat capacity versus
temperature. It was interpreted as indicating a phase transition to
a magnetically ordered phase. An important characteristic of the
low-dimensional magnets is the absence of the long range order in
models with a continuous symmetry at any finite
temperature\cite{mermin66}. There is  also a broad maximum in the
plot of the heat capacity vs temperature, characteristic of low
dimensional systems. This is known as the Schottky peak.

In this paper, theoretical and numerical results are reported for
the low-temperature behavior of the heat capacity in low dimensional
spin systems. Theoretically, by keeping only two lowest states, the
system is mapped to the well known two-level-system (TLS) model. In this case,
the heat capacity is found exactly as a function of the energy gap
and the temperature. It is shown that the position of 
the Schottky heat capacity peak, $T_{M}$, and the energy gap 
behaves in the same way as a function of
the control parameter. In
Section II, the mapping to the TLS model is explained and the theoretical
results are presented. In Section III, the results of the low
temperature Lanczos calculations are presented. Numerically, $T_{M}$
as a function of the control parameter is computed for "the
alternating spin-1/2 chains in a magnetic field $h$" and "the 1D
Heisenberg Hamiltonian with a staggered magnetic field $h_s$".
Finally, the summary and conclusions are presented in Section IV
%%%%%%%%%%%%%%%%%%%%%%%%%%%%%%%%%%%%%%%%%%%%%%%%%%%%%%%%%%%%%%%%%%%%%%%%%%
%%%%%%%%%%%%%%%%%%%%%%%%%%     Section II     %%%%%%%%%%%%%%%%%%%%%%%%%%%%
%%%%%%%%%%%%%%%%%%%%%%%%%%%%%%%%%%%%%%%%%%%%%%%%%%%%%%%%%%%%%%%%%%%%%%%%%%
\section{Low temperature limit heat capacity: Two Level System approach}

In this section we discuss a theoretical approach to find the
effect of the energy gap on the heat capacity of the quantum  spin
systems. Spectrum energy of a quantum system may be gapful or
gapless. Since, we are going to study the sign of the gap on the heat
capacity, gapful systems are considered.   At very low temperature
we can consider only lowest energy levels. If we keep only  two
lowest energy states, the system maps to the two level system (TLS)
model\cite{TLS-model}. We have assumed that the energy gap of this TLS is $g$ where
$E_1=E_0+g$, $E_0$ and $E_1$ are the ground and first exited
states respectively.

\begin{figure}
\centerline{\includegraphics[width=8cm,angle=0]{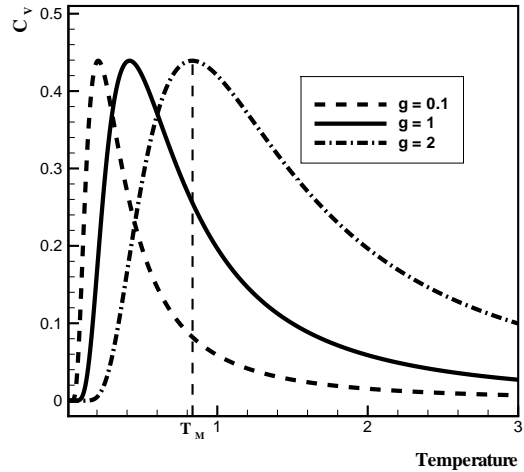}}
\caption{Temperature dependence of the  heat capacity of TLS
(Eq.~\ref{eq4}) for different energy gap values (g). $T_M$ shows
the position  of the Schottky heat capacity peak for the
system with  $g=2$.
 }
 \label{figa}
\end{figure}

Our purpose is to determine the behavior of the heat capacity for
the quantum spin systems. The heat capacity is expressed by the
following relation
\begin{equation}
C_v =\frac{1}{K_BT^2}\left[ \frac{\partial ^{2}\ln Z}{\partial
\beta^{2}}\right].
\label{eq1}
\end{equation}
 where $Z$ is the partition function  denoted by
\begin{equation}
Z \; = \; {\rm Tr}\left\{ e^{-\beta H}\right\},
\label{eq2}
\end{equation}
Here
 $\beta =
\frac{1}{ K_BT}$, $K_B$ is the Boltzmann constant and $T$ is
temperature. We have assumed that $K_B=1$. At very low temperature
limit (TLS limit), we can write
\begin{equation}
Z \simeq e^{-\beta E_0}+e^{-\beta E_1}=e^{-\beta E_0}(1+e^{-\beta
g}),
\label{eq3}
\end{equation}
And finally from the above equation one can shows
\begin{equation}
C_v =\frac{x^2}{\cosh ^2(x)},
\label{eq4}
\end{equation}
where we defined $x=\frac{g}{2T}$. This result predict a Schottky
like peak of the heat capacity behavior versus the $x$ variation.
The position of the Schottky  peak takes place at
$x_M\simeq 1.2$, where we  have: $\tanh(x_M)=1/x_M$.  This result
shows explicitly an upward increase of the heat capacity versus the
magnetic field for
 $x<x_M $ and monotonic decrease
for $x>x_M$.  We have plotted the thermal behavior of
Eq.~\ref{eq4} in Fig.~\ref{figa}. Here, $x_M$ corresponds to the
position of the Schottky peak, $T_M$, in a
constant gap value. It is clear that by increasing the gap value,
$|g|$, the peak approach to higher temperature and inversely.
Therefore we can conclude that width of the energy gap may affect
the position of the Schottky heat capacity peak. At very
low temperature regime, all gapped thermodynamic systems
($N\rightarrow \infty$, $N=$ number of the spins) can be mapped to
the above TLS model.

Up to now we did not consider  any degeneracy. In the general case
both first two energy levels (TLS) have a degeneracy. 
It is well known  for two level system
(TLS) that the account of degeneracy leads to change the effective gap and consequently to
change the $T_M$
 for Schottky anomaly of heat capacity\cite{Wunderlich}.
In peresent of degeneracy by a little of manipulation,   one can shows
\begin{equation}
C_v =\frac{x^2}{\cosh ^2(x+x')},
\label{eq7}
\end{equation}
which is very similar to Eq.~\ref{eq4}. Here we 
denoted $x'=\frac{g_{s}}{2T}$, and $g_{s}=T\ln {\frac{d_2}{d_1}}$.
Where  $d_1$ and $d_2$ have been considered 
for the order of the degeneracy of the
ground state and first exited state of the system respectively. 
As we mentioned it before this result predict a Schottky like peak
in the heat capacity behavior. The Schottky heat capacity  peak takes place at
$x_M$ which satisfied  $\tanh(x_M + x'_M)=1/x_M$,
 where $x'_M=\frac{g_{s}}{2T_M}$. If we assumed that
$d_1<d_2$ then $g_s<0$ and therefore the generalized gap ($g+g_{s}$) will be
smaller than the gap. Thus the Schottky  peak moves to the higher
values of the gap in respect to the case $g_s=0$ (without
degeneracy). In same manner for the case: $d_2<d_1$ we have $g_s>0$,
therefore the generalized gap will be larger than the gap. Thus the
Schottky  peak moves to the lowest values of the gap in respect to
the case $g_s=0$.
%%%%%%%%%%%%%%%%%%%%%%%%%%%%%%%%%%%%%%%%%
%%%%%%%%%%%%%%%%%%%%%%%%%%%%%%%%%%%%%%%%%%

%%%%%%%%%%%%%%%%%%%%%%%%%%%%%%%%%%%%%%%%%%%%%%%%%%%%%%%%%%%%%%%%%%%%%%%%%%
%%%%%%%%%%%%%%%%%%%%%%%%%%     Section III    %%%%%%%%%%%%%%%%%%%%%%%%%%%%
%%%%%%%%%%%%%%%%%%%%%%%%%%%%%%%%%%%%%%%%%%%%%%%%%%%%%%%%%%%%%%%%%%%%%%%%%%
\section{Numerical results}

In recent years, numerical methods have been extensively developed
and applied to quantum many-body problems. Most frequently used
numerical method for these
 problems is the exact diagonalization of small systems employing the Lanczos
 technique\cite{lanczos50}. The exact diagonalization of small correlated systems
 does not have any restrictions on the model. The deficiency of the method is
 in the relative smallness of system sizes. So far the method has been essentially
 restricted to the evaluation of the $T=0$ static and dynamical quantities, i.e.,
 properties of the ground state.

Jakli\~c et.al. introduced a method for the evaluation of finite-temperature properties,
based on the Lanczos diagonalization technique for small systems\cite{jaklic94}.
This method, is avoid the calculation of all eigenfunctions of the system. Instead,
they introduced the procedure where the sampling over all states is reduced to a
random partial sampling, while only approximate ground state and excited state wave
functions, generated by the Lanczos technique, are used for the evaluation of matrix
elements. The size limitations of the method are effectively comparable to those
encountered in the Lanczos-type diagonalization technique applied to the ground
state calculations.

In following, we  present our numerical results on the heat capacity
of the several 1D spin-1/2 models which are obtained by the method of
Jakli\~c.

%%%%%%%%%%%%%%%%%%%%%%%%%%%%%%%%%%%%%%%%%%%%%%%%%%%%%%%%%%%%%%%%%%%%%%%%%%%%%%%%
%%%%%%%%%%%%%%%%%%%%%%%%%%%%         SubSdection I         %%%%%%%%%%%%%%%%%%%%%
%%%%%%%%%%%%%%%%%%%%%%%%%%%%%%%%%%%%%%%%%%%%%%%%%%%%%%%%%%%%%%%%%%%%%%%%%%%%%%%%
\subsection{Alternating  Heisenberg  Spin-1/2  Chains in a Transverse Magnetic
Field }

In this section we consider the alternating spin-1/2 chains
in a magnetic field. Since, the antiferromagnetic-ferromagnetic
 (AF-F) chains have a gap in the spin excitation spectrum, they
  reveal extremely rich quantum behavior in the presence of the
  magnetic field.

The ground state phase diagram of the AF-F alternating chain in a
magnetic field is studied by the numerical diagonalization and the
finite-size scaling based on the conformal field
theory\cite{Sakai95b}. It is shown that the magnetic state is
gapless and described by the Luttinger liquid phase. It is also
found that the magnetic state is characterized by the algebraic
decay of the spin correlation functions. Recently, Yamamoto et.al
described the magnetic properties of the model in a magnetic field
in terms of the spinless fermions and the spin
waves\cite{Yamamoto05}. They employed the Jordan-Wigner
transformation and treated the fermionic Hamiltonian within the
Hartree-Fock approximation. They have also implemented the modified
spin wave theory to calculate the thermodynamic functions as the
heat capacity and the magnetic susceptibility.
%%%%%%%%%%%%%%%%%%%%%%%%%%%%%%%%%%%%%%%%%
\begin{figure}
\centerline{\includegraphics[width=8cm,angle=0]{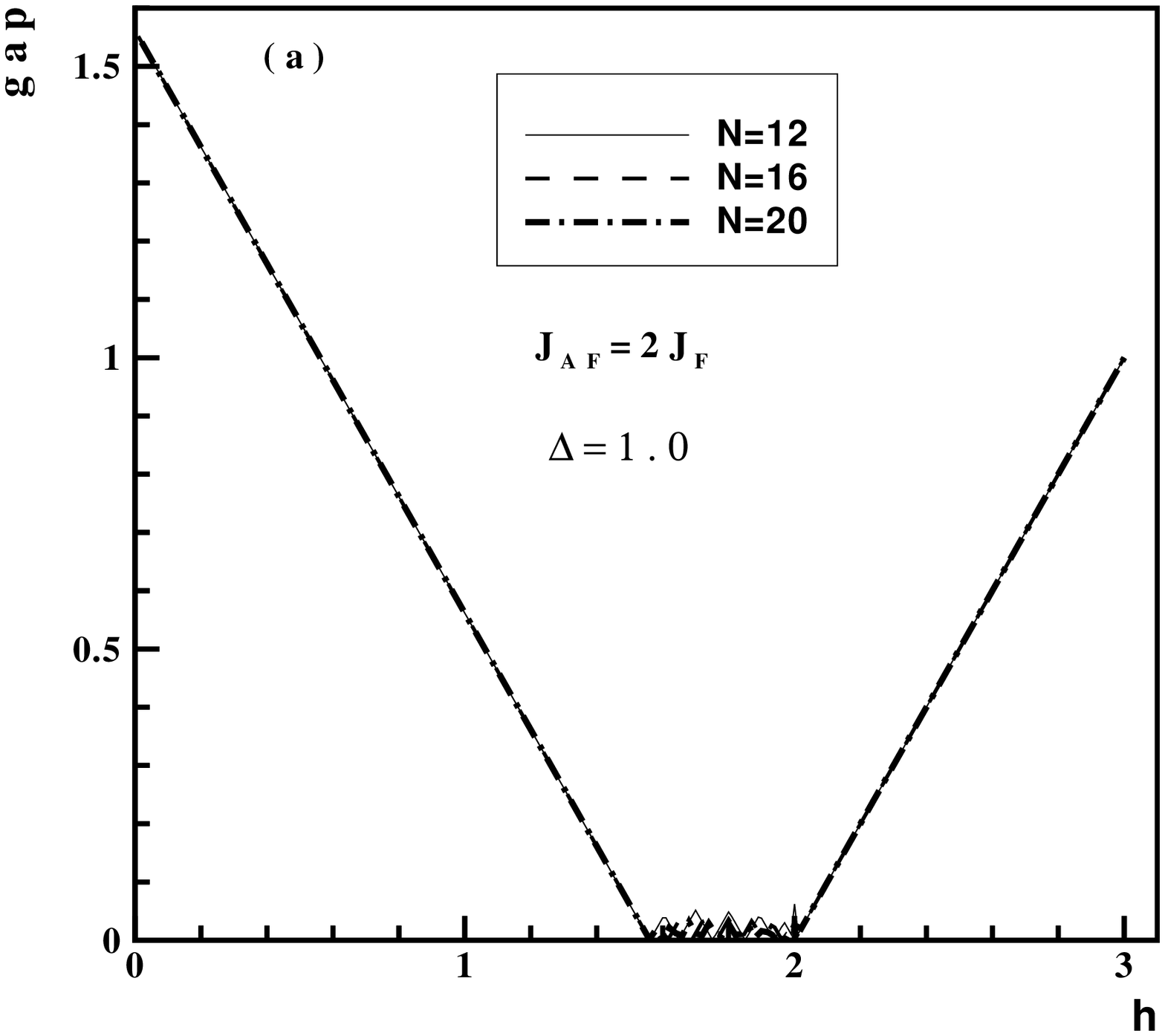}}
\centerline{\includegraphics[width=8cm,angle=0]{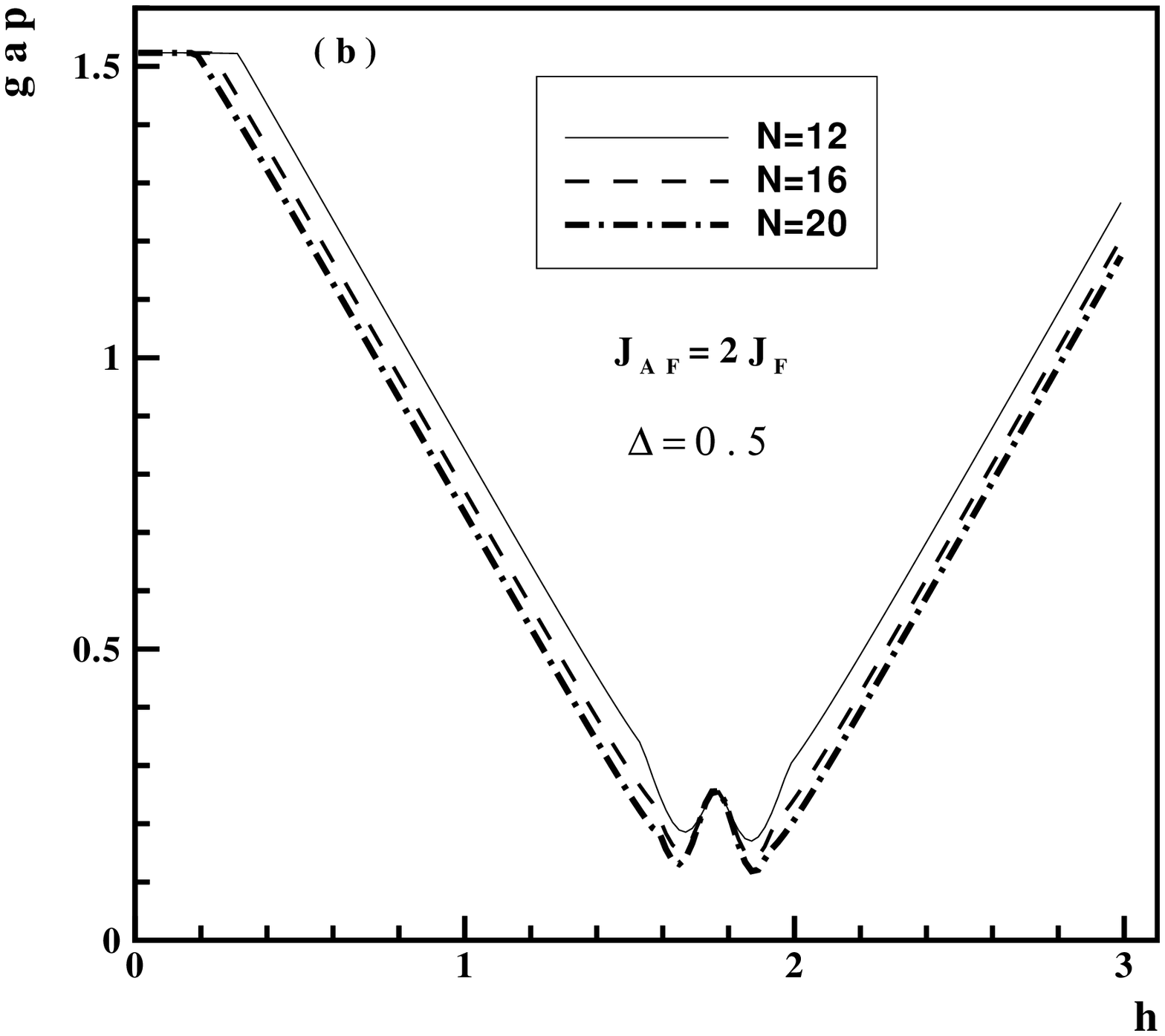}}
\caption{a. The excitation gap of a spin-1/2 AF-F chain for
isotropic case $\Delta=1.0$ in the uniform  magnetic field for
different size number ($N=12,16, 20$).  b. The excitation gap of a
spin-1/2 AF-F chain in the uniform transverse magnetic field with
anisotropic ferromagnetic coupling $\Delta=0.5$ , for different
size number ($N=12,16, 20$).
 } \label{fig1}
\end{figure}
%%%%%%%%%%%%%%%%%%%%%%%%%%%%%%%%%%%%%%%%%%
%%%%%%%%%%%%%%%%%%%%%%%%%%%%%%%%%%%%%%%%%
\begin{figure}
\centerline{\includegraphics[width=8cm,angle=0]{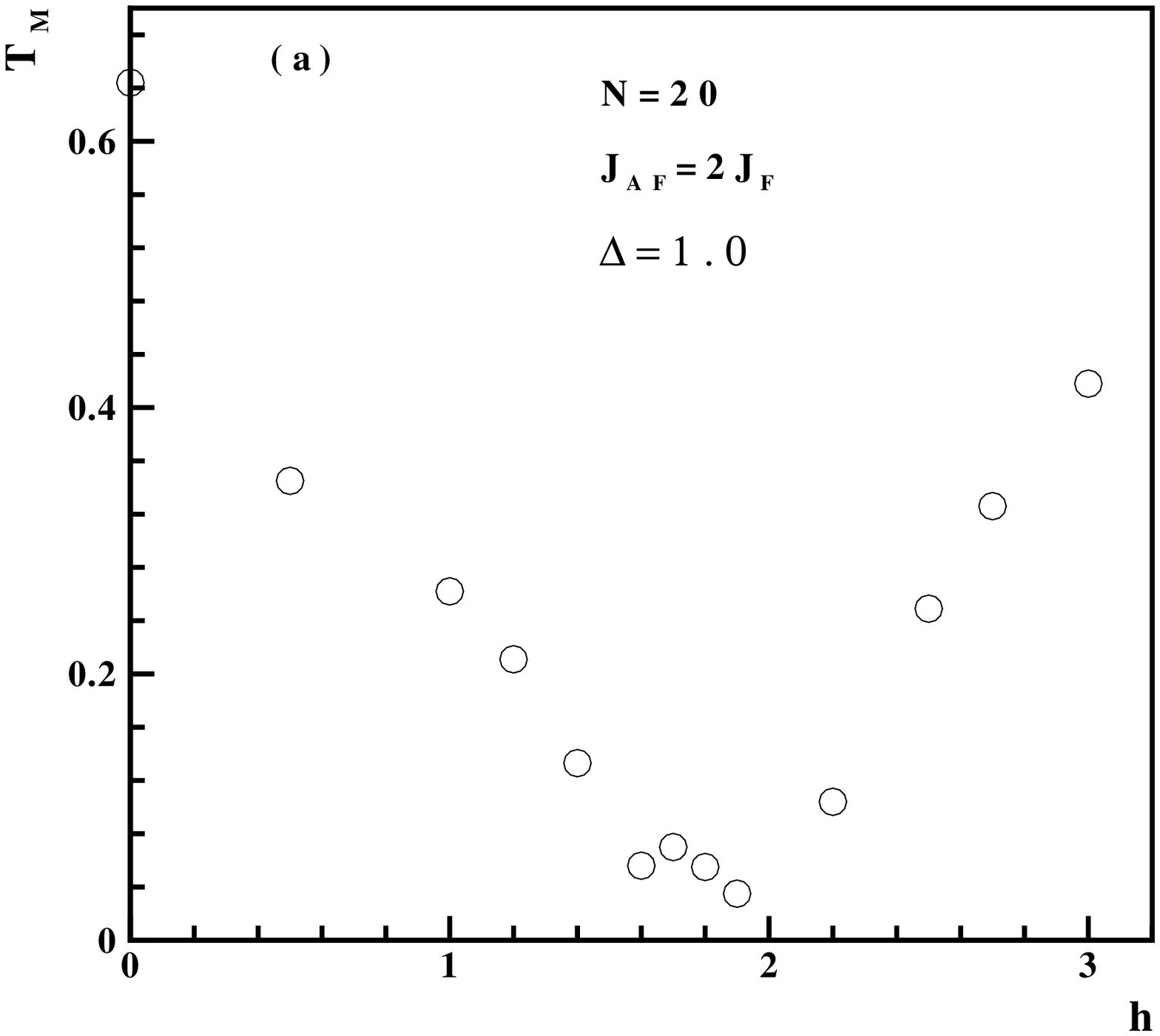}}
\centerline{\includegraphics[width=7.65cm,angle=0]{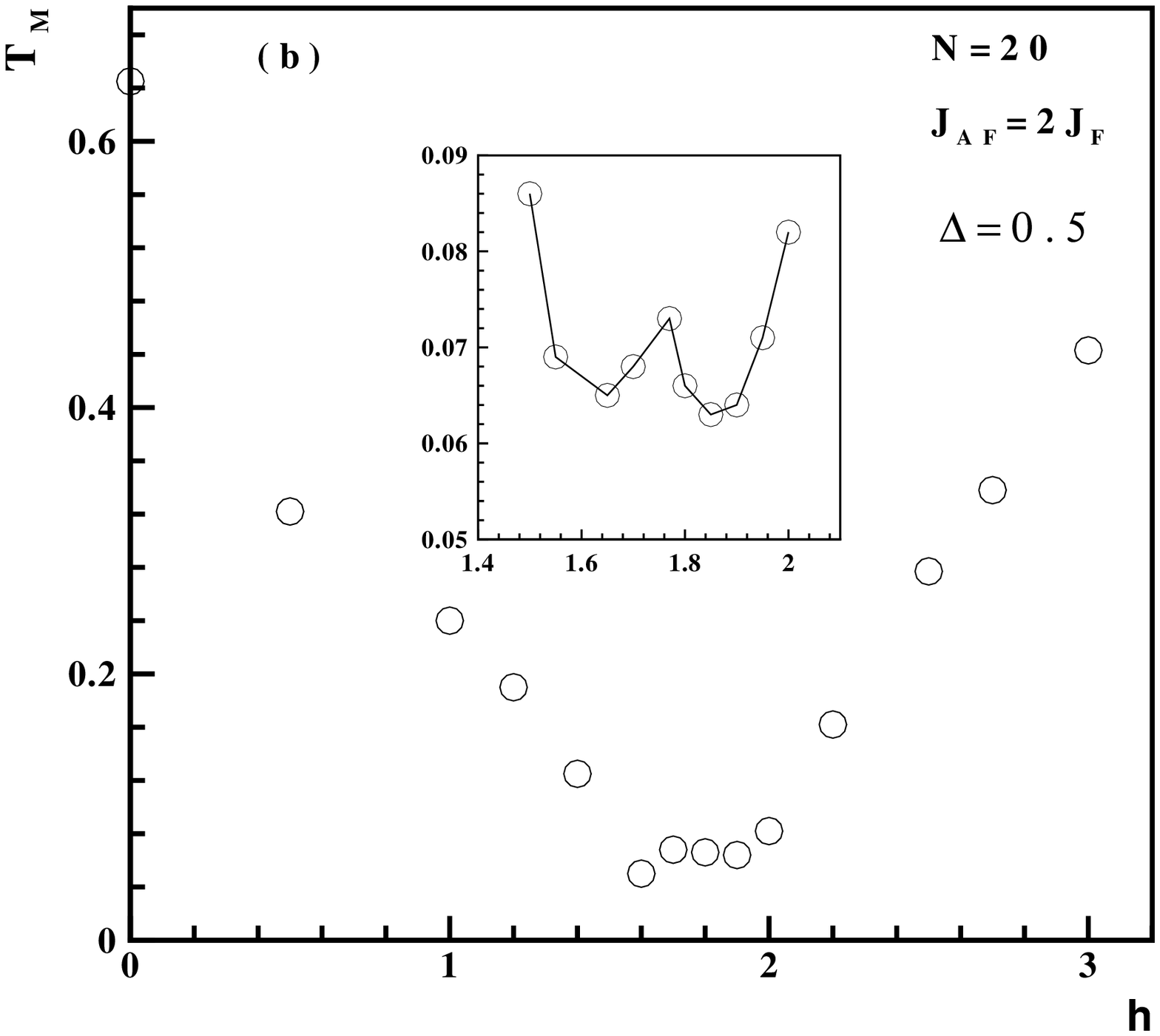}}
\caption{a. The Schottky heat capacity peak value,  $T_M$, versus
the transverse magnetic field, for   a spin-1/2 AF-F chain, and
isotropic case $\Delta=1.0$ in the uniform transverse magnetic
field for system size:  $N=20$.  
b. The  Schottky heat capacity peak value, $T_M$,
as a function of the magnetic field, for a
spin-1/2 AF-F chain in the uniform transverse magnetic field with
anisotropic ferromagnetic coupling: $\Delta=0.5$, and size number:
$N=20$. The inset shows the same quantity in the intermediate
magnetic field region, $h_{c_{1}}<h<h_{c_{2}}$.
 }
 \label{fig2}
\end{figure}
%%%%%%%%%%%%%%%%%%%%%%%%%%%%%%%%%%%%%%%%%%
More recently, using numerical Lanczos method, the effect of
an uniform transverse magnetic field on the ground
state phase diagram of a spin-1/2 AF-F chain with anisotropic
ferromagnetic coupling is studied \cite{saeed4}. The
Hamiltonian of the model under consideration on a periodic chain
of $N$ sites is given by
\bea
 H & =&  J_{AF}\sum_{j=1}^{N/2}[S_{2j-1}^x S_{2j}^x + S_{2j-1}^yS_{2j}^y
+ S_{2j-1}^zS_{2j}^z]
\no& -& J_F\sum_{j=1}^{N/2} [S_{2j}^xS_{2j+1}^x +
S_{2j}^yS_{2j+1}^y
 +\Delta S_{2j}^zS_{2j+1}^z]
\no& +& h\sum_{j=1}^N
 S_{j}^{x}.
 \label{Hamiltonian}
\eea
Where $S_{j}^{x, y, z}$ are spin-1/2 operators on the
$j$-th site. $J_{F}$ and $J_{AF}$ denote the ferromagnetic and
antiferromagnetic couplings respectively. The limiting case of
isotropic ferromagnetic coupling corresponds to $\Delta=1$ and $h$
is the transverse magnetic field.
To explore the nature of the excitation spectrum, we use the modified Lanczos method to
diagonalize numerically finite chains ($N=12,16, 20, 24$).
The energies of the few lowest eigenstates were obtained for the
chains with periodic boundary conditions. First, we have computed the three lowest energy
eigenvalues of $N=12, 16, 20$ chain with $J_{AF}=2 J_{F}$ and
different values of the anisotropy parameter $\Delta$.

In Fig.~\ref{fig1}a we have plotted results of calculations
for the isotropic case $\Delta=1.0$. The excitation gap is
determined\cite{saeed4}
 in the system as the difference between the first excited state and the ground state.
As it is clearly seen from this figure in the case of zero magnetic
field the spectrum of the model is gapped. For $h\neq 0$ the gap
decreases linearly with $h$ and vanishes at the critical field,
$h_{c_{1}}=1.55\pm0.01$. This is the first level crossing between
the
 ground state energy and the first excited state. To get an accurate estimate of $h_{c_{1}}$
we have obtained the first level crossing for system sizes of $N=12,
14, ..., 24$. The finite size behavior of these values lead us to
$h_{c_{1}}=1.55\pm0.01$ for $N\longrightarrow \infty$. The spectrum
remains gapless for $h_{c_{1}}<h<h_{c_{2}}$ and becomes once again
gapped for $h>h_{c_{2}}=2.0$. With increasing field, for
$h>h_{c_{2}}$ the gap increases linearly with $h$. In the region
$h_{c_{1}}<h<h_{c_{2}}$ we also observe numerous additional level
crossing between the lowest eigenstates. These level crossing lead
to incommensurate effects that manifest themselves in the
oscillatory behavior of the spin correlation functions. All
crossings disappear at $h>h_{c_{2}}$ and the correlation functions
do not contain oscillatory terms in this region of the phase
diagram.

In marked contrast with the isotropic case, the similar analysis
of the few lowest levels for an anisotropic AF-F chain in the
presence of a transverse magnetic field reveal a principally
different behavior. The gap as a function of the transverse
magnetic field $h$ has been computed for the anisotropy parameter
$\Delta=0.5$ and different chain lengths $N=12, 16, 20$. In
Fig.\ref{fig1}b we have plotted results of these calculations.
 As it is seen from the figure, the excitation spectrum in this case is gapful
 except at the two critical fields $h_{c_{1}}=1.64\pm0.01$ and $h_{c_{2}}=1.88\pm0.01$\cite{saeed4}.
  We have employed the phenomenological renormalization group
(PRG) method\cite{Barber}  to determine these critical 
fields ($h_{c_1}$ and  $h_{c_2}$). The PRG equation is
%***********************************************************
\begin{eqnarray}
(N+4)gap(N+4,h')=Ngap(N,h),
\label{prg}
\end{eqnarray}
%******************************************************** 
 where $gap(N,h)=E_1(N,h) - E_0(N,h)$ is the energy gap value for
 chain length $N$ in a magnetic field $h$.
 At the critical point, $N(E_1 - E_0)$  should be size independent
for large enough systems in which the contribution
from irrelevant operators is negligible. Thus,
we accurately determined the  critical points by the
PRG method. We defined $h_c(N,N+4)$ as the $N$-dependent fixed point of 
Eq.~\ref{prg}, and it is extrapolated to the thermodynamic limit 
in order to estimate $h_{c}$.
At the critical point $h=h'=h_{c}$, therefore, the curves
of  $N(E_1 - E_0)$ vs $h$ for sizes $N$ and $N + 4$ cross at certain values
 $h_{c_1}(N,N +4)$ and $h_{c_2}(N,N +
4)$ (’finite-size critical points’). The thermodynamic critical points 
($h_{c_1}$ and  $h_{c_2}$)
 are obtained by
appropriately extrapolating $h_{c_1}(N,N +4)$ or $h_{c_2}(N,N + 4)$ 
to $N\rightarrow \infty$.
 
 In the region $h_{c_{1}}<h<h_{c_{2}}$ the spin gap, which appears at $h>h_{c_{1}}$,
 first increases vs external field and after passing a maximum decreases to vanish at
$h_{c_{2}}$. At $h>h_{c_{2}}$ the gap once again opens and, for a
sufficiently large transverse field becomes proportional to $h$.
%%%%%%%%%%%%%%%%%%%%%%%%%%%%%%%%%%%%%%%%%
\begin{figure}
\centerline{\includegraphics[width=8cm,angle=0]{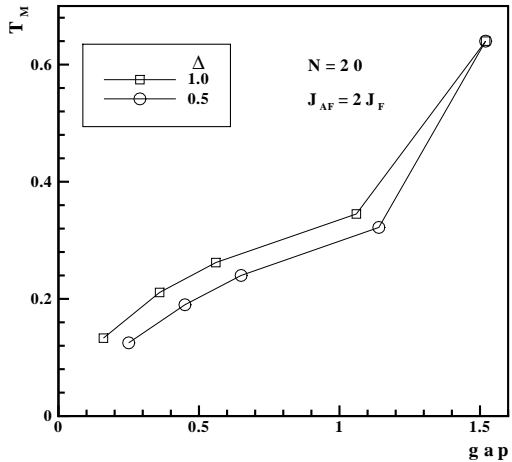}}
\caption{The Schottky heat capacity peak value,  $T_M$, versus
the energy gap for a
spin-1/2 AF-F chain with
anisotropic ferromagnetic couplings: $\Delta=1.0$ and $\Delta=0.5$, and size number:
$N=20$. 
 } \label{fig3}
\end{figure}
%%%%%%%%%%%%%%%%%%%%%%%%%%%%%%%%%%%%%%%%%%
To study the finite-temperature properties of the model, we have
used the Jakli\~c formalism. We have computed a hundred lowest
eigenvalues of the energies. We have considered different values of
the transverse magnetic field, $h$, and anisotropy parameter
$\Delta$. Therefore using these hundred eigen-energies,  we have
computed the heat capacity as a function of the temperature ($T$).
The  position of 
the Schottky heat capacity peak, $T_{M}$ is determined as
$T_M$. In Fig.~\ref{fig2}a we have plotted $T_M$ as a function of
the magnetic field $h$. To arrive at this plot, we have considered
$J_{AF}=2J_F$, $\Delta=1$ and $N=20$.
As it is clearly seen from this figure, the position of 
the Schottky heat capacity peak, $T_{M}$, $T_M$, decreases by increasing the
magnetic field up to the critical field, $h_{c_1}$. In the
intermediate region of the magnetic filed,
$h_{c_{1}}<h<h_{c_{2}}$, $T_M$ is independent of magnetic field.
The anomaly behavior is the result of the finite size effects.
 With increasing the magnetic
field, for $h>h_{c_{2}}$, $T_M$ increases almost linearly.

It is surprising which the  behavior of $T_M$ versus the magnetic
field is in complete agreement with the gap behavior respect to
the magnetic filed. Thus we  conclude that the energy gap sign on
the position of 
the Schottky heat capacity peak. To confirm
our idea, we have plotted $T_M$ vs the transverse magnetic field
for $J_{AF}=2J_F$, the anisotropy parameter $\Delta=0.5$ and
$N=20$ in
 Fig.~\ref{fig2}b.
 As we can see from this figure for low transverse
 magnetic field, $h<h_{c_{1}}$,  the behavior of
 $T_M$ is the same as the isotropic case.
But in the intermediate region, $h_{c_{1}}<h<h_{c_{2}}$,  $T_M$
first increases vs transverse field and after passing a maximum,
decreases up to $h_{c_{2}}$ (see the inset of the
Fig.~\ref{fig2}b). For the higher transverse magnetic field,
$h>h_{c_{2}}$, the position of 
the Schottky heat capacity peak
increases as a previous one. This behavior of the $T_M$ is in
complete agreement with the effect of the transverse magnetic
field on the energy gap (Fig.~\ref{fig1}b).

Finally, we have plotted $T_M$ versus the energy gap for $N=20$, $J_{AF}=2J_F$ and different values of the anisotropy parameter $\Delta=0.5, 1.0$. For convenience, the numerical results of the region $h<h_{c_{1}}$ are showed. It is clearly seen, that the possition of the Schottky heat capacity peak $T_M$, increases by increasing the energy gap of the system. This is in well agreement with results obtained within the two-level model.

%%%%%%%%%%%%%%%%%%%%%%%%%%%%%%%%%%%%%%%%%%%%%%%%%%%%%%%%%%%%%%%%%%%%%%%%%%%%%%%%
%%%%%%%%%%%%%%%%%%%%%%%%%%%%         SubSdection II        %%%%%%%%%%%%%%%%%%%%%
%%%%%%%%%%%%%%%%%%%%%%%%%%%%%%%%%%%%%%%%%%%%%%%%%%%%%%%%%%%%%%%%%%%%%%%%%%%%%%%%
\subsection{The 1D AF-Heisenberg model in a staggered field }

The general feature developed for the alternating spin-1/2 chains in a
 magnetic field can be applied to the 1D Heisenberg Hamiltonian with a
staggered magnetic field $h_s$,
%***********************************************************
\begin{eqnarray}
H=J\sum_{i=1}^{N}[\overrightarrow{S}_{i}.\overrightarrow{S}_{i+1}+h_{s}
(-1)^{i}S_{i}^{z}].  \label{hamiltoni3}
\end{eqnarray}
%***********************************************************
%%%%%%%%%%%%%%%%%%%%%%%%%%%%%%%%%%%%%%%%%
\begin{figure}
\centerline{\includegraphics[width=8cm,angle=0]{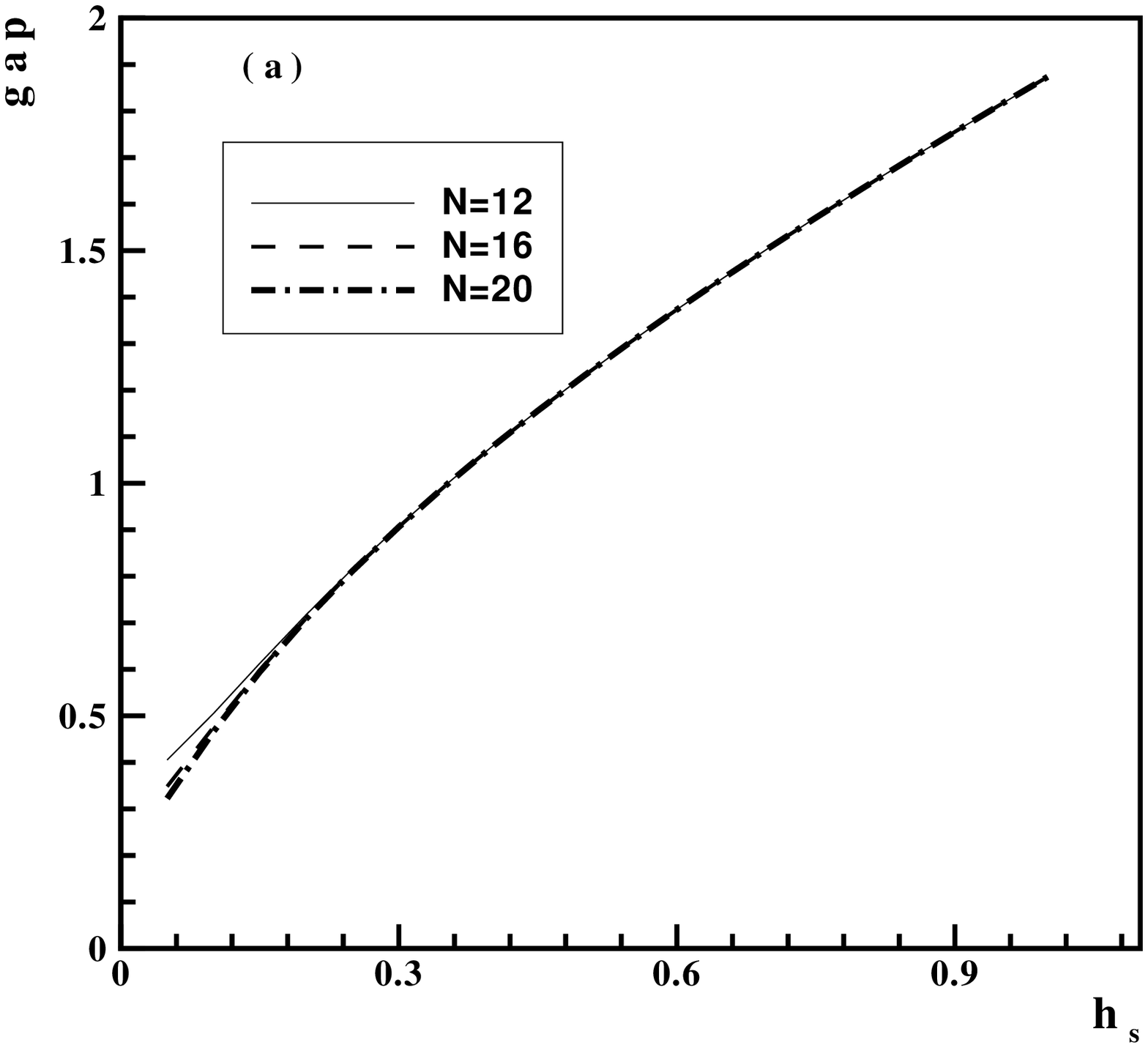}}
\centerline{\includegraphics[width=8cm,angle=0]{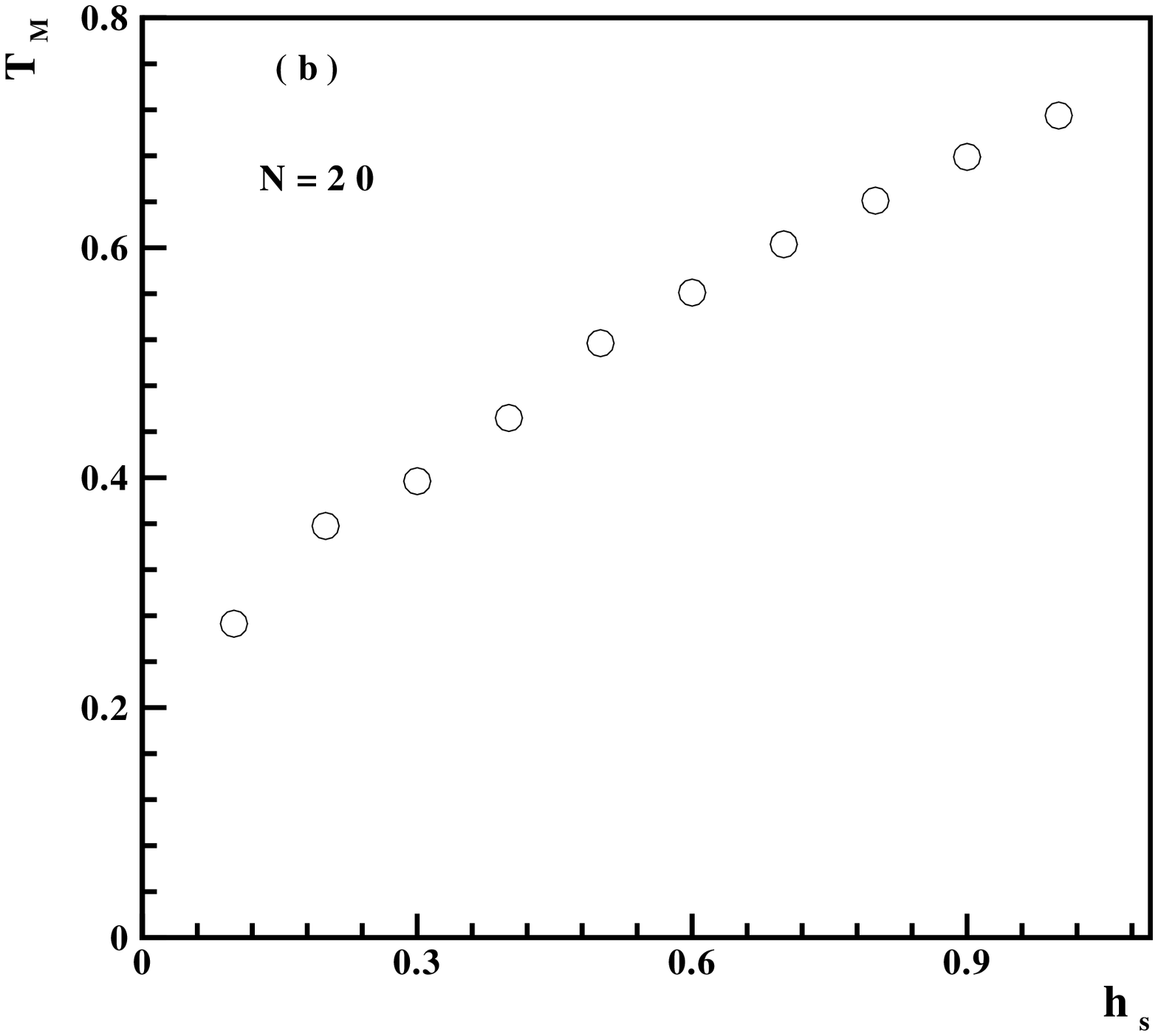}}
\caption{a. The excitation gap of the 1D AF-Heisenberg model in a
staggered field  as a function of the magnetic field for different
size number ($N=12,16, 20$). 
b. The position of 
the Schottky heat capacity peak, $T_{M}$, for the 1D AF-Heisenberg 
model in a staggered field
versus the  staggered magnetic field with  size number: $N=20$.
 }
 \label{fig3}
\end{figure}
%%%%%%%%%%%%%%%%%%%%%%%%%%%%%%%%%%%%%%%%%%
It is expected\cite{oshikawa2, wang, sato, alcaraz1} that the
staggered field induces an excitation gap in the $S=\frac{1}{2}$
AF-Heisenberg chain, which should be otherwise gapless. The
excitation gap caused by the staggered field is indeed found in
the real magnets\cite{dender, kohgi, feyerherm}. In the absence of
the staggered field ($h_{s}=0$), the eigenspectra is exactly
solvable. In the case of the staggered magnetic field
($h_{s}\neq0$), the integrability is lost. The staggered magnetic
field produces an antiferromagnetic ordered (Neel order)
ground-state.

To examine the effect of the staggered magnetic field on the
energy gap,
 we have implemented the modified Lanczos algorithm for finite-size chains
  $N=12, 16, 20$ using periodic boundary conditions. The energy gap
  is determined as the difference between the first excited
state and the ground state\cite{saeed1}, and calculated for different chain lengths and staggered fields $h_{s}$.
   The energy gap is determined as the difference between the first excited
state and the ground state\cite{saeed1}.

We have plotted, in Fig.~\ref{fig3}a, the energy gap versus the staggered magnetic field $h_{s}$.
The results have been plotted for different chain sizes $N=12, 16, 20$. It can be seen, that the spectrum is
gapless in the absence of the staggered magnetic field ($h_{s}=0$). The application of a staggered magnetic field,
induces a gap in the spectrum of the model. With increasing the field, for $h_{s}>0$, the energy gap increases with
$h_{s}$.

Applying the Jakli\~c method we have computed a hundred lowest
energies for different values of the staggered magnetic field. The
heat capacity is computed as a function of the temperature. In
Fig.~\ref{fig3}b, the position of 
the Schottky anomaly heat capacity peak, $T_{M}$
  is plotted versus $h_{s}$ for the chain
size $N=20$. As it is seen from the figure, $T_{M}$ in this case
increases by increasing the staggered magnetic field. This result is
in good agreement with the idea that the gap sign on the position of 
the Schottky heat capacity peak, $T_{M}$.

%%%%%%%%%%%%%%%%%%%%%%%%%%%%%%%%%%%%%%%%%%%%%%%%%%%%%%%%%%%%%%%%%%%%%%%%%%%%%%%%
%%%%%%%%%%%%%%%%%%%%%%%%%%%%     Summary and discussion    %%%%%%%%%%%%%%%%%%%%%
%%%%%%%%%%%%%%%%%%%%%%%%%%%%%%%%%%%%%%%%%%%%%%%%%%%%%%%%%%%%%%%%%%%%%%%%%%%%%%%%
\section{Summary and discussion}
%%%%%%%%%%%%%%%%%%%%%%%%%%%%%%%%%%%%%%%%%%%%%%%%%%%%%%%%%%%%%%%%%%%%%%%%%%%%%%%%
Low temperature behavior of the heat capacity of the
low-dimensional spin systems is studied using theoretical and
numerical approaches. Theoretically, the system is mapped to the
well known two-level-system (TLS) model. In this case, the heat capacity is
found exactly as a function of the energy gap and the temperature.
The position of the Schottky heat capacity peak, $T_{M}$, is
determined. It is shown that $T_{M}$ as a function of the control
parameter behaves in the same way as the energy gap versus the
control parameter. This shows that the gap has an influence on the
position of the Schottky heat capacity peak.

Numerically, the finite temperature Lanczos method is applied. The
Lanczos method is implemented to obtain a hundred of lowest excited
state energies. This formalism is applied to two model chains up to
$N=24$ in length. First, the alternating spin-1/2 chains in a
magnetic field are considered. Since, the
antiferromagnetic-ferromagnetic
 (AF-F) chains have a gap in the spin excitation spectrum, they
  reveal extremely rich quantum behavior in the presence of the
  magnetic field. The energy gap and heat capacity are computed for
  both isotropic and anisotropic cases. The numerical results are
  computed for different values of the external magnetic field. It
  is shown, in complete agreement with the theoretical results,
  the field-dependence of the $T_{M}$ and the energy
  gap are the same. Finally, the 1D Heisenberg model with a
staggered magnetic field $h_s$ is investigated. It is shown that the
staggered field induces an excitation gap in the $S=\frac{1}{2}$
AF-Heisenberg chain, which should be otherwise gapless. Using the
above numerical procedure, the position of 
the Schottky heat capacity peak is computed. It is confirm, that
the energy gap sign on the Schottky heat capacity peak.

%%%%%%%%%%%%%%%%%%%%%%%%%%%%%%%%%%%%%%%%%%%%%%%%%%%%%%%%%%%%%%%%%%%%%
\section{Acknowledgments}
We would like to thank J. Abouie
for insightful comments and  stimulating discussions. We are also
grateful to B. Farnudi and M. Aliee for reading our
manuscript carefully and appreciate their useful comments.
%%%%%%%%%%%%%%%%%%%%%%%%%%%%%%%%%%%%%%%%%%%%%%%%%%%%%%%%%%%%%%%%%%
%%%%%%%%%%%%%%%%%%%%%%%%%%%%%%%%%%%%%%%%%%%%%%%%%%%%%%%%%%%%%%%%%%%%%%%%%%%%%%%%
%%%%%%%%%%%%%%%%%%%%%%%%%%%%%      References
%%%%%%%%%%%%%%%%%%%%%%%%%%%%%%%%
%%%%%%%%%%%%%%%%%%%%%%%%%%%%%%%%%%%%%%%%%%%%%%%%%%%%%%%%%%%%%%%%%%%%%%%%%%%%%%%%

\end{document}